\documentclass[11pt]{article}

\let\ssection=\section
\renewcommand{\section}{\setcounter{equation}{0}\ssection}

\def\p{{\partial}}

\newcommand{\br}{{\bf r}}
\newcommand{\bq}{{\bf q}}
\newcommand{\bk}{{\bf k}}

\newcommand{\half}{{\scriptstyle{\frac{1}{2}}}}
\def\det{{\rm det}}
\def\bA{{\bf A}}

\begin{document}

\setlength{\baselineskip}{16pt}

\title{Hamiltonian aspects of Bogoliubov quasiparticles} 

\author{
Z.~Horv\'ath\footnote{e-mail: zalanh@ludens.elte.hu}
\\
Institute for Theoretical Physics, E\"otv\"os
University\\
H-1117 BUDAPEST (Hungary)
\\
P.~A.~Horv\'athy\footnote{e-mail: horvathy@lmpt.univ-tours.fr}
\\
Laboratoire de Math\'ematiques et de Physique Th\'eorique\\
Universit\'e de Tours\\
F-37 200 TOURS (France)
\\
L. Martina\footnote{e-mail: Luigi.Martina@le.infn.it}
\\
INFN and Dipartimento di Fisica, Universit\`a di Lecce
\\
I-73 100 LECCE (Italy)
}

\date{\today}

\maketitle

\begin{abstract}
The Bogoliubov particle considered in
[cond-mat/0507125] admits, contrarily to the claim of the authors, an interesting Hamiltonian structure.
\end{abstract}

\noindent
\texttt{cond-mat/0511099}.
\vskip5mm

In \cite{NewNiu}  Zhang et al.  derive,
from the small deviation (Bogoliubov) equation for 
a superfluid condensate, a semiclassical model, where position and momentum satisfy 
\begin{equation}
\dot{\br}=\frac{\p h}{\p\bk},
\qquad
\dot{\bk}=-\frac{\p h}{\p\br},
\label{caneq}
\end{equation}
where $h$, the  energy [$\omega$ in their \# (12)],
 is a rather complicated function of $\br$ and $\bk$, that also involves
a Berry vector potential $\bA=\bA(\bq,\br)$, where  
$ 
\bq=\bk-\bA.
$ 
Eliminating $\bk$ in favor of $\bq$ and $\br$ transforms
 (\ref{caneq}) into a 
more complicated form, (\ref{eqmot}) below,
which, Zhang et al. claim, would {\it no longer be Hamiltonian}
and would lead to a {\it a violation of Liouville's theorem} on the conservation of the phase space volume. These statements stem from a
misinterpretation of Hamiltonian mechanics \cite{correct}.
As we show below, the
system admits in fact an interesting Hamiltonian structure,
and the Liouville theorem is  not violated~: one simply has to use the correct phase-space volume.

Eqns. (\ref{caneq}) are Hamiltonian~: they
correspond to the Hamiltonian $h=h(\bk,\br)$ and to the
canonical Poisson brackets. 
 But to be Hamiltonian is an intrinsic property 
of the system that can  {\it not} be lost by a mere change of
variables~: the error is
that the Authors of \cite{NewNiu}
transform the Hamiltonian 
but seem to forget about transforming simultaneously the Poisson structure.

The situation is conveniently explained working with
the symplectic structure 
$\Omega=\half\Omega_{\alpha\beta}d\xi^\alpha\wedge d\xi^\beta$,
where $\xi^\alpha$ are coordinates on the phase space, and
$\Omega_{\alpha\beta}$ is the inverse of the Poisson
matrix $\Pi^{\alpha\beta}=\{\xi^\alpha,\xi^\beta\}$
 \cite{HamDyn}. 
 The equations of motion  are 
$\Omega_{\alpha\beta}\dot{\xi}^\alpha=\p_{\beta}h$.
In canonical coordinates, $\Omega=d\bk\wedge d\br$ and we get (\ref{caneq}).
The change of variables $(\bk,\br)\to(\bq,\br)$  takes this into
\begin{eqnarray}
\Omega=\half C_{ij}dq^i\wedge dr^j-\half F_{ij}dr^i\wedge dr^j,
\qquad
C_{ij}=\delta_{ij}-\frac{\p A_i}{\p q^j},
\qquad
F_{ij}=\frac{\p A_j}{\p r^i}-\frac{\p A_i}{\p r^j}.
\label{newsymp}
\end{eqnarray}
$C$ plays the role of an effective mass matrix; 
the equations of motion become Eq. (13) of \cite{NewNiu},
\begin{eqnarray}
C\dot{\br}=\frac{\p h}{\p\bq}
\qquad 
-C\dot{\bq}+F\dot{\br}=\frac{\p h}{\p\br} .
\label{eqmot}
\end{eqnarray}

If $\bA$ was a function of $\br$ alone, we would have $C={\bf 1}$, and
for $h(\bq,\br)=\bq^2/2m+V(\br)$
we would recover  
a particle with unit charge in an electromagnetic field.
The $\bq$-dependent case is more interesting, though.
Assuming that the system is regular, 
det$(\Omega_{\alpha\beta})=\left[{\rm det\,}(C_{ij})\right]^2\neq0$, the 
matrix $C=(C_{ij})$ will be invertible, and the
 Poisson commutation relations become
\begin{equation}
\begin{array}{lll}
\{q^i,q^j\}=\left(C^{-1}FC^{-1}\right)_{ij},
\qquad 
\{q^i,r^j\}=-\left(C^{-1}\right)_{ij},
\qquad 
\{r^i,r^j\}=0.
\end{array}
\label{Poisson}
\end{equation}
The first of these generalizes the usual momentum-momentum relations
in a magnetic field;
the second modifies the ``Heisenberg" commmutation relations of momentum and position. The coordinates (Poisson-) commute.
The equations of motion (\ref{eqmot}) are also obtained from the Hamiltonian framework as
$
\dot{x}^i=\{x^i,H\},
\,
\dot{q}^i=\{q^i,H\}
$
with $H(\bq,\br)=h\big(\bk(\bq,\br),\br\big)$.

When
$ {\rm det\,}\Omega=
{\rm det\,}C^2\equiv {\rm det}\left({\bf 1}-{\p\bA}/{\p\bq}\right)^2=0,
$ 
the system becomes singular. This case, although 
spurious for the model of \cite{NewNiu},
is nevertheless interesting. The equations of motion (\ref{eqmot}) can
indeed remain consistent when $\p_{\bq}h=0$ (which fixes $\bq$,
which are no more dynamical), provided the motion follows a  {\it generalized Hall law},
\begin{equation}
\epsilon_{ijk}B^k\dot{r}^j=-E_i
\end{equation}
where $E_i=-\p h/\p r_i$ and $B_i=\half\epsilon_{ijk}F_{jk}$ are generalized electric and magnetic fields.
The trajectories are defined by the vanishing of the ``Lorentz''
force.
In this case, the Hamiltonian structure (\ref {Poisson}) blows up; 
Hamiltonian reduction \cite{FaJa} would yield a lower
dimensional system. 

Turning to the Liouville theorem, let us emphasise
that the volume element of the phase space can {\it only}
be defined through the symplectic form \cite{HamDyn}, 
\begin{equation}
dV=\sqrt{\det(\Omega_{\alpha\beta})}\prod_{\alpha}d\xi^\alpha.
\label{volel}
\end{equation}
The pre-factor det$\,(\bf 1-\p\bA/\p\bq)$ ``discovered'' by Zhang et al. is precisely the square-root
of the determinant of the symplectic matrix. It
is also the square-root of the Jacobian, $J$, of the transformation
from canonical to arbitrary
variables, and is hence always present whenever the change of variables
is non-canonical.
Now the Liouville theorem \cite{HamDyn} 
 says that  the symplectic volume element
 is invariant w.r.t. the Hamiltonian flow. 

In the singular case  the Liouville volume form
(\ref{volel}) becomes degenerate.

Analogous problems were studied before in \cite{DH, correct}.


\end{document}